\documentstyle[pre,aps,multicol,epsfig]{revtex}

\begin{document}
\draft

\title{Counterion correlations and attraction between like-charged macromolecules}
\author{\bf Alexandre Diehl$^{1}$, Humberto A. Carmona$^{2}$, and Yan Levin$^{3}$}
\address{\it $^1$Departamento de F{\'\i}sica, 
Universidade Federal do Cear{\'a}\\
Caixa Postal 6030, CEP 60455-760, Fortaleza, CE, Brazil\\
$^2$Departamento de F\'\i sica e Qu\'\i mica, Universidade Estadual do Cear\'a\\    
CEP 60740-000 Fortaleza, Cear\'a, Brazil\\  
$^3$Instituto de F{\'\i}sica, Universidade Federal do Rio Grande do Sul\\
Caixa Postal 15051, CEP 91501-970, Porto Alegre, RS, Brazil
}
\date{\today}
\maketitle

\begin{abstract}

A simple model is presented for the appearance of attraction between
two like charged polyions inside a polyelectrolyte solution. The polyions
are modeled as rigid cylinders in a continuum dielectric solvent. The strong 
electrostatic interaction between the polyions and the counterions results 
in counterion condensation. If the two polyions are sufficiently close to 
each other their layers of condensed counterions can become correlated 
resulting in attraction between the macromolecules. To explore the counterion 
induced attraction we calculate the correlation functions for the condensed 
counterions. It is found that the correlations are of very short range. For 
the parameters specific to the double stranded DNA, the correlations and the 
attraction appear only when the surface-to-surface separation is less 
than $7\;\mbox{\AA}$.

\end{abstract}
\pacs{PACS number(s): 61.20.Qg, 87.15.Aa, 87.15.-v}

\begin{multicols}{2}

\narrowtext

\section{Introduction}
\label{introd}

In the last few years a new phenomenon has attracted attention of the community 
of soft condensed matter physicists --- appearance of attraction between like 
charged macromolecules in solutions containing multivalent ions. The problem 
is particularly fascinating because it contradicts our well established intuition 
that like charged entities should repel~\cite{Bloomfield91,Podgornik94,Israelachvili85}. 
The fundamental point, however, is that the electrolyte solutions are intrinsically 
complex systems for which many body interactions play a fundamental role.  

The attraction between like charged macromolecules is important for many biological 
systems. One particularly striking example is provided by the condensation of DNA 
by multivalent ions such as $\mbox{Mn}^{2+}$, $\mbox{Cd}^{2+}$ and various 
polyamines~\cite{Rau92,Duguid95,Knoll98}. This condensation provides an answer to 
the long standing puzzle of how a highly charged macromolecule, such as the DNA, 
can be confined to a small volume of viral head or nuclear zone in procaryotic cell. 
Evidently, the multivalent ions serve as a glue which keeps the otherwise repelling 
like-charged monomers in close proximity~\cite{ras98}. In eukaryotic cells, the 
cytosol is traversed by a network of microtubules and microfilaments --- rigid chains 
of highly charged protein (F-actin) ---  which in spite of large negative charge 
agglomerate to form filaments of cytoskeleton~\cite{Tang96}. The actin fibers are also 
an important part of the muscle tissue, providing a rail track for the motion of 
molecular motor myosin.  

Although the nature of attraction between like charged macromolecules is still not 
fully understood, it seems clear that the attractive force is mediated by the 
multivalent counterions~\cite{Ray94,Ha97,Ha00,Shklovsk199,Shklovsk299,Perel99,Nguyen00}. 
A strong electrostatic attraction between the polyions and the oppositely charged 
multivalent counterions produces a sheath of counterions around each macromolecule. 
The condensed counterions can become highly correlated resulting in an overall 
attraction. It is important to note that the complex formed by a polyion and 
its associated counterions does not need to be neutral for the attraction to arise. 
Under some conditions the correlation induced attraction can overcome the monopolar 
repulsion coming from the net charge of the complexes.   

Recently a simple model was presented to account for the attraction between two 
lines of charges~\cite{Levin99,Arenzon99,Arenzon00}. Each line had $Z$ discrete 
uniformly spaced monomers of charge $-q$, and $n \leq Z/\alpha$ condensed counterions 
of charge $\alpha q$ free to move along the rod. The net charge of such a 
polyion-counterion complex is $Q_{\scriptscriptstyle\rm eff}=-(Z-\alpha n)q  \leq 0$. 
Nevertheless, 
it was found that if $n>Z/2\alpha$ and $\alpha \geq 2$, 
at sufficiently short distances, the two 
like-charged rods would attract~\cite{Arenzon99}. It was argued that the attraction 
resulted from the correlations between the condensed counterions and reached 
maximum at zero temperature. If $n<Z/2\alpha$ the force was always found to be repulsive. 

Clearly, a one dimensional line of charge is a dramatic oversimplification of 
the physical reality. If we are interested in studying the correlation induced 
forces between real macromolecules their finite radius must be taken into 
account~\cite{Niels97,Solis99,Lowen00}. Thus, a much more realistic model of a 
polyion is a cylinder with a uniformly charged backbone~\cite{Niels97,Solis99} or 
with an intrinsic charge pattern~\cite{Leikin99,Lowen00} as, e.g., the helix structure 
of DNA molecule. Furthermore, the condensed counterions do not move along the 
line, but on the surface of the cylinder. Unfortunately, these extended models are 
much harder to study analytically. 

In this paper we explore the effects of finite polyion diameter on the electrostatic 
interactions between the two polyions using Monte Carlo simulations. We find that 
the finite diameter and the associated angular degrees of freedom of condensed 
counterions significantly modify the nature of attraction. Thus, although there 
is still a minimum charge which must be neutralized by the counterions in order 
for the attraction to appear, this fraction is no longer equal to $50\%$ as 
was the case for the line of charge model. We find that the critical fraction 
depends on the valence of counterions and is less than $50\%$ for $\alpha \geq 2$. 
For monovalent counterions no attraction is found. The crystalline structure of 
the condensed counterions, as first suggested by simulations of 
Gronbech-Jensen {\it et al.}~\cite{Niels97} and 
Refs.~\cite{Shklovsk199,Levin99,Arenzon00,Solis99}, is also not very obvious. 
In particular we find very similar distributions of condensed counterions in 
the regime of attractive and repulsive interactions.  

The structure of this paper is as follows. The model and the method of calculation 
are described in section~\ref{model}. In section~\ref{results}, we present 
the results of the simulations. The conclusions are summarized in section~\ref{summary}. 

\begin{figure}[t]
\begin{center}
\epsfxsize=0.4\textwidth
\epsfbox{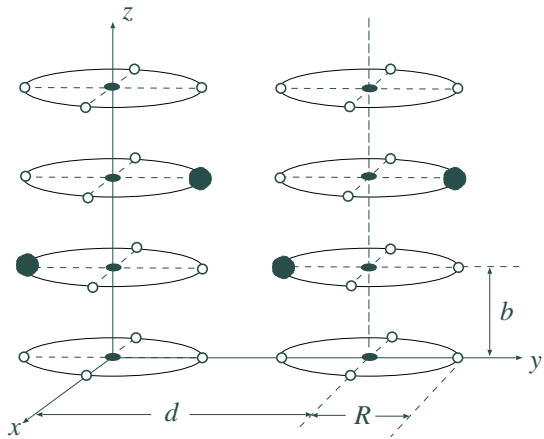}
\end{center}
\caption{Schematic  depiction of the model: two polyions of $Z$ 
negative charges (small solid circles at the centers of rings), with 
radius $R$ and $n$ condensed counterions (large solid circles) each, 
are separated by distance $d$. The counterions are free to move along 
the $Zm$ positions (open circles) fixed on the rings located around 
each of the $Z$ monomers. In this sketch, $Z=4$, $n=2$ and $m=4$.}
\label{modelfig}
\end{figure}

\section{Model and method}
\label{model}

\subsection{The model}

\begin{figure}[t]
\begin{center}
\epsfxsize=0.4\textwidth
\epsfbox{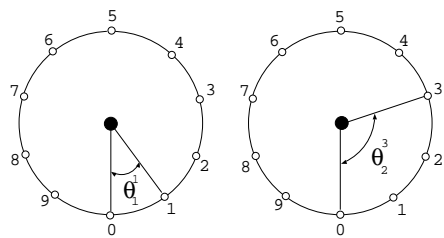}
\end{center}
\caption{The cross sectional view of the adjacent rings on the two polyions. 
The sites are labeled with integers $j=0, 1, \ldots ,m-1$ in the 
counter-clockwise direction. The angle between two consecutive sites is 
$\Delta \theta = 2\pi/m$. Here we show the angles $\theta_1^1$ of site 
$i=1$ on the polyion 1, and $\theta_3^2$ of site $i=3$ on the polyion 2.}
\label{rings}
\end{figure}

The DNA model considered here is an extension of the one proposed earlier 
by Arenzon, Stilck and Levin~\cite{Levin99,Arenzon99}. A similar model has been 
recently discussed by Solis and Olvera de la Cruz~\cite{Solis99}. The polyions 
are treated as parallel rigid cylinders of radius $R$ and $Z$ ionized groups, 
each of charge $-q$, uniformly spaced --- with separation $b$ --- along the 
principle axis, Fig.~\ref{modelfig}. Besides the fixed monomers, each polyion 
has $n \leq Z/\alpha$ condensed counterions with valence $\alpha$ and 
charge $\alpha q$, which are constrained to move on the surface of the cylinder. 
To locate a condensed counterion it is necessary to provide its longitudinal 
position, $z$ ($0\leq z <Z$), and the transversal angle, 
$\theta$ ($0\leq \theta \leq 2\pi$). To simplify the calculations, the angular 
and the longitudinal degrees of freedom are discretized, see Fig.~\ref{modelfig}. 
The surface of the cylinder is subdivided into $Z$ parallel rings with a charged 
monomer at the center of each ring. Each ring has $m$ sites available to the 
condensed counterions, see Figs.~\ref{modelfig} and \ref{rings}. The hardcore 
repulsion between the particles requires that a site is occupied by at most one 
condensed counterion. The two polyions are parallel, with the intermolecular 
space treated as a uniform medium of dielectric constant $\epsilon$.  

We introduce occupation variables $n_i^k$ for the two polyions, so that 
$i=1, 2, \ldots, Z(m+1)$ and $k=1,2$. Thus, $n_{i}^k = 1$ if the $i$'th site of the 
$k$'th polyion is occupied by a particle of valence $\alpha_i=\{-1,\: \alpha\}$ 
(negative core charge or counterion of valence $\alpha$, respectively), 
otherwise $n_{i}^k = 0$. Note that the core charge is always ``occupied'', while 
the counterions are free to move between the $Zm$ ring sites of each 
polyion. The Hamiltonian for the interaction between the two polyions is, 
\begin{equation}
\label{hamiltonian}
\beta  H = \xi \sum_{k,l}\sum_{i,j}
\frac{\alpha_i^k n_i^k\;\alpha_j^l n_j^l}
{|\mbox{\boldmath$r$}_i^k-\mbox{\boldmath$r$}_j^l|}\:,
\end{equation} 
with the $i \neq j$. All lengths are measured in units of $b$ 
(for DNA, $b=1.7\:$\AA). The dimensionless quantity, $\xi =\beta q^2/b\epsilon$, 
is the Manning parameter, which for DNA is $\xi = 4.17$. The partition function 
is obtained by tracing over all the possible values of $n_i^k$ consistent 
with the constraint of fixed number of counterions per polyion, 
\begin{equation}
\label{partition}
{\cal Q} = \sum_{\{n_i^k\}} \exp (-\beta H)\:.
\end{equation}

Clearly, this is a very crude model of the interaction between two macromolecules 
in a polyelectrolyte solution. The molecular nature of the solvent is ignored. 
Also the number of condensed counterions is fixed instead of being dependent on 
the separation between the particles. Nevertheless, we believe that this 
simple model can provide some useful insights for the mechanism of attraction 
in real polyelectrolyte solutions. 

\begin{figure}[t]
\vspace*{-1cm}
\begin{center}
\epsfxsize=0.35\textwidth
\epsfbox{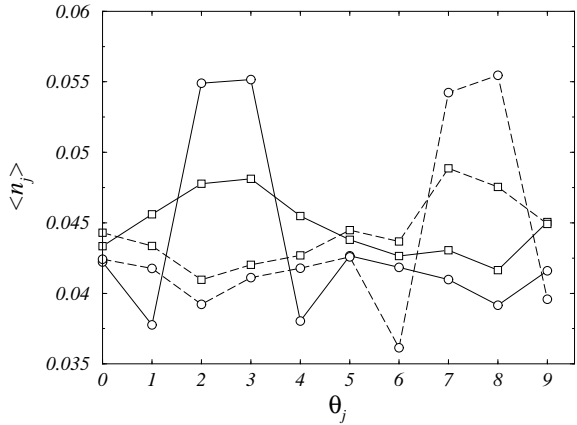}
\end{center}
\vspace*{-1.cm}
\caption{Mean site occupation for the adjacent rings on the two polyions located at 
$z=9$. Each polyion has $Z=20$ and $n=7$. The distance between the centers of 
rings is $d/b=32.8$ (squares) and $d/b=16.65$ (circles). The solid line corresponds 
to the sites on the first polyion and the dashed lines are for the second polyion. 
Note the almost perfect symmetry between the two macromolecules.}
\label{z20n7_ocup}
\end{figure}

\subsection{The observables}

We are interested in statistical averages of observables such as the energy and 
the force between the two polyions. Furthermore, to understand the nature of 
the interaction between the two macromolecules it is essential to study the 
correlations between the condensed counterions on the two polyions. 

The force is obtained from the partition function Eq.~(\ref{partition}), 
$b \beta \mbox{\boldmath$F$} = -\mbox{\boldmath$\nabla$}(\ln \cal Q)$. From symmetry, 
only the $y$-component is different from zero.

For finite macromolecules the symmetry between the two polyions can not be 
broken~\cite{Arenzon00}. Hence it is impossible to produce a true crystalline order 
in a finite system at non-zero temperature. Since within our simplified model 
the two polyions have exactly the same number of condensed counterions, the 
average angular counterion distribution $\langle n_i^k(z,\theta^k_i)\rangle$ must 
be symmetric with respect to the mid-plane $y=d/2$, see Fig.~\ref{modelfig}. The 
angle $\theta_i^k$ labels the site $i$ on polyion $k$, see Fig.~\ref{rings}. 
Thus, $n_3^2(z,\theta_3^2)$ denotes the occupation variable for the site 3 on the 
ring $z$ located on polyion 2, with an angle of $3(2\pi/m)$. Indeed, 
Fig.~\ref{z20n7_ocup} shows that the density profiles are completely symmetric 
(up to fluctuations). In spite of this symmetry it is possible for the counterions 
on the two polyions to become highly correlated. Clearly, the strength of these 
correlations will depend on the product $\xi \alpha^2$ and the separation between 
the two macromolecules. Considering Fig.~\ref{rings}, it is evident that if the 
site $2$ on the first polyion is occupied, the likelihood of occupation of the 
site $8$ on the second polyion will be reduced. 
 
To explore the nature of electrostatic correlations, we define a counterion-hole 
correlation function between the adjacent rings on the two polyions,   
\begin{eqnarray}
\label{correl}
c^{12}_{zij} &=& \langle n_i^1(z,\theta^1_i) \left[ 1-n_j^2(z,\theta^2_j)\right] \rangle \nonumber\\
&&- \langle n_i^1(z,\theta^1_i)\rangle \: \langle \left[1-n_j^2(z,\theta^2_j)\right] \rangle \:.
\end{eqnarray}
Here $\langle \cdots \rangle$ denotes the ensemble average. This function should 
be non-zero when sites on the two polyions are correlated, that is if one is occupied 
by a condensed counterion there is an increased probability of the second being empty.

\subsection{Simulations}

To calculate the force between the two polyions, we have performed a standard 
Monte-Carlo (MC) simulation with the usual Metropolis algorithm~\cite{Allen}. 
First, one counterion on polyion 1 is randomly chosen and displaced to a vacant 
position on the {\it same} polyion. This move is accepted or rejected according 
to the standard detailed balance criterion~\cite{Allen}. We do not permit 
exchange of particles between the polyions. Next, the same is done for polyion 2. 
In one Monte-Carlo step (MCS) all $2n$ condensed counterions on the two polyions 
are permitted to attempt a move. 

The long-ranged nature of the Coulomb interaction requires evaluation of all the pair 
interactions in Eq.~(\ref{hamiltonian}) at every MCS. Due to the limited computational 
power available to us, we have confined our attention to relatively small systems 
with $Z<100$ and $m=10$. We have checked, however, that for $m=10$ the force has 
already reached the continuum limit and did not vary further with increase of 
$m$. Also we note that the ``thermodynamic limit'' is reached reasonably quickly, 
so that there is a good collapse of data already for $Z > 50$, see Fig.~\ref{z20_force}. 
2000 MCS served to equilibrate the system, after which 500 samples were used to 
calculate the basic observables, namely, the mean force and energy. To obtain the 
correlation functions, 5000 samples were used with 5000 MCS for equilibration.  

\section{Results and Discussion}
\label{results}

The simulations were performed for $\xi = 4.17$ and $R/b=8.2$, relevant for 
DNA. For monovalent counterions the simulation results indicate that the 
force is purely repulsive. This is in complete agreement with the 
experiments~\cite{Bloomfield91}, which do not find any indication of 
DNA condensation for monovalent counterions.

\begin{figure}[t]
\vspace*{-1cm}
\begin{center}
\epsfxsize=0.35\textwidth
\epsfbox{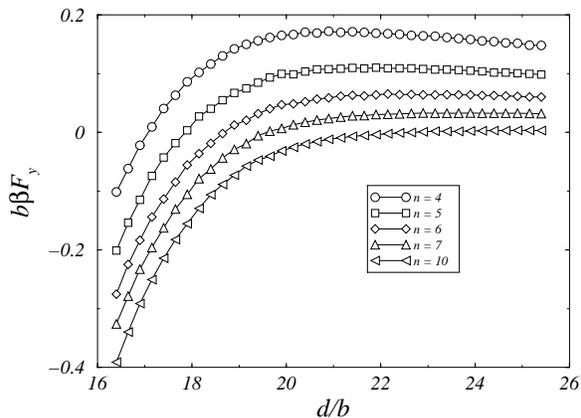}
\end{center}
\vspace*{-1.cm}
\caption{Mean force in $y$-direction {\it versus} distance $d/b$ between the two DNA 
molecules $R/b=8.2$, $Z=20$ and $\xi =4.17$. The symbols indicate the number 
$n$ of divalent ($\alpha =2$) condensed counterions.}
\label{z20_force}
\end{figure}
 
For divalent counterions the force between the two complexes can becomes negative, 
indicating appearance of an effective attraction, Fig.~\ref{z20_force}. The range 
of attraction is larger than was found for the one dimensional line of 
charge model, Ref.~\cite{Arenzon99}. 

\begin{figure}[h]
\vspace*{-1cm}
\begin{center}
\epsfxsize=0.35\textwidth
\epsfbox{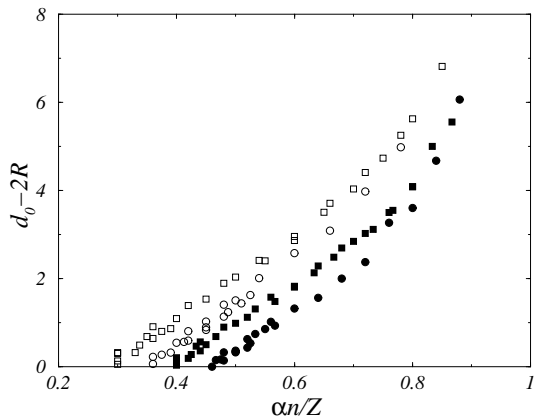}
\end{center}
\vspace*{-.5cm}
\caption{The surface-to-surface separation below which the force becomes attractive, 
as a function of the number $n$ of condensed counterions, for valences 
$\alpha = 2$ (full) and 3 (open), $\xi = 2.283$ (circles) and $4.17$ (squares), 
with $Z$ ranging from 50 to 100.}
\label{d_0}
\end{figure}

Within the Manning theory~\cite{man69} $88\%$ of the DNA's charge is neutralized 
by the divalent counterions. However, there are indications that even a larger 
fraction of DNA's charge can become neutralized by the multivalent ions if the 
counterion correlations are taken into account~\cite{Perel99}. In this case the 
interaction is purely attractive, with the range of about $d/b \approx 20$ 
or $34\;\mbox{\AA}$ 
($7\;\mbox{\AA}$ surface-to-surface), Fig.~\ref{z20_force}. 

\begin{figure}[t]
\begin{center}
\epsfxsize=0.4\textwidth
\epsfbox{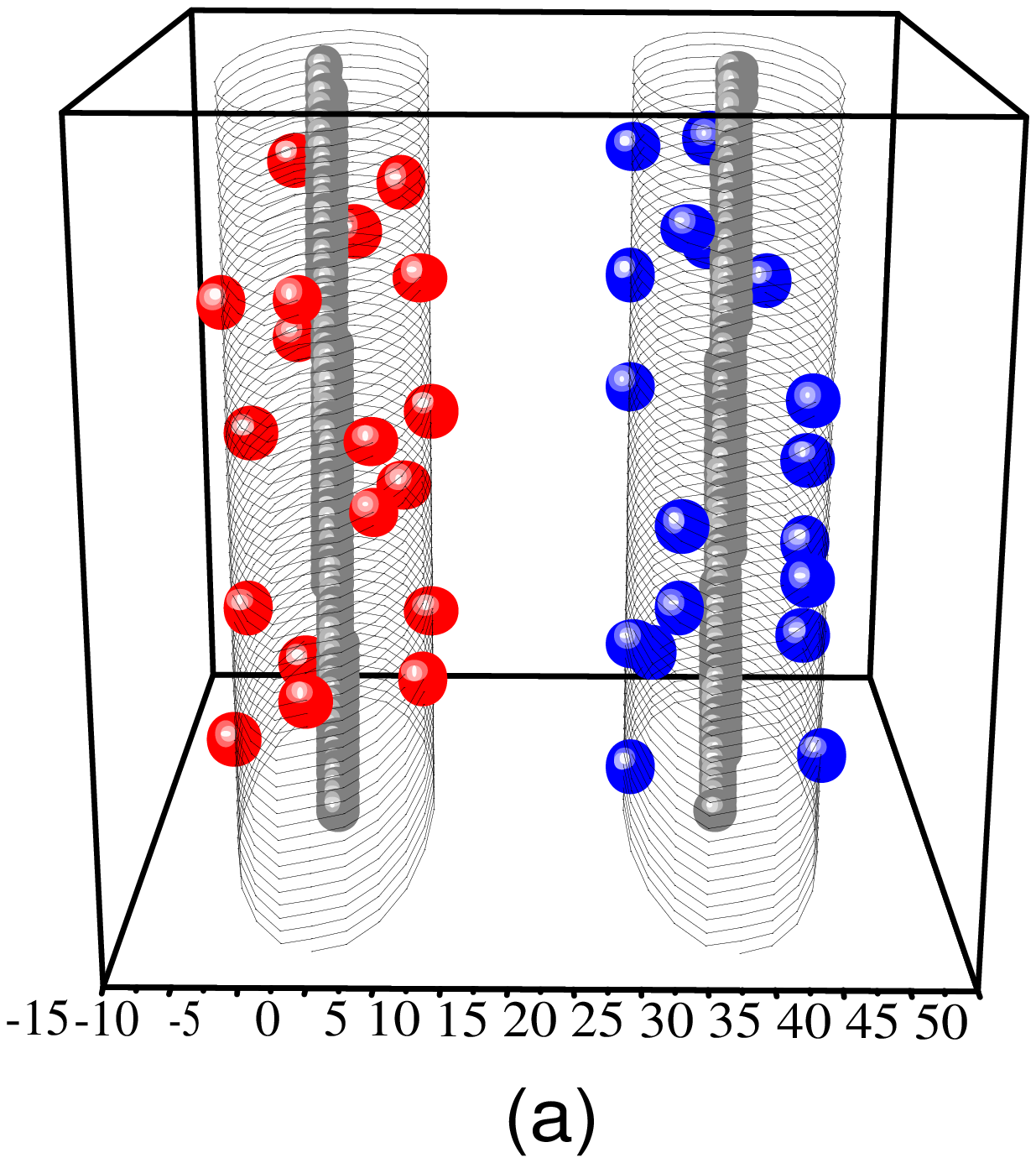}
\end{center}
\end{figure}

\begin{figure}[t]
\begin{center}
\epsfxsize=0.4\textwidth
\epsfbox{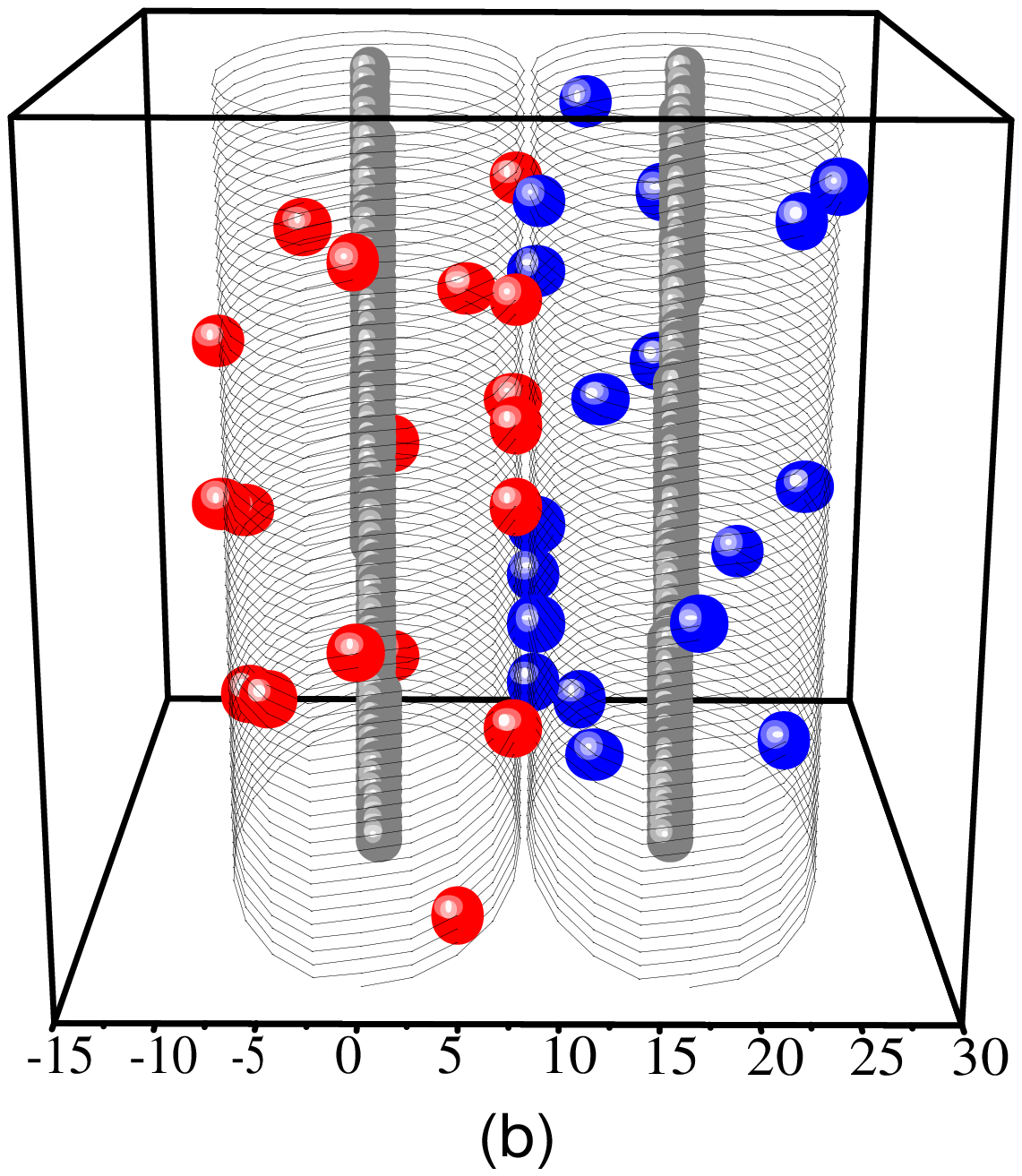}
\end{center}
\caption{Snapshots of two equilibrium configurations for (a) $d/b=32.8$ and 
(b) $d/b=16.8$, for two polyions with $Z=60$ and $n=18$.}
\label{snaps}
\end{figure}

A minimum number of condensed counterions is necessary for attraction 
to appear. In Fig.~\ref{d_0} we present the surface-to-surface separation, 
$(d_0 - 2R)$, below which the force between the two complexes becomes negative 
(attractive), as a function of the number of multivalent counterions. For the 
case of DNA with divalent counterions $\alpha =2$, the attraction appears 
only if $40\%$ of the core charge is neutralized. For $\alpha = 3$ this fraction 
decrease to $30\%$. Furthermore, decrease in the value of the Manning parameter, 
$\xi$, increases the minimum number of condensed counterions necessary for 
the attraction to appear. This is fully consistent with the fact that the 
attraction is mediated by the correlations between the condensed counterions. 
Since raise in temperature tends to disorganize the system, the state of 
highest correlation between the condensed counterions corresponds 
to $T=0$ or $\xi=\infty$. 
 
The surface-to-surface distance at which the attraction first appears tends to zero 
as the number of condensed counterions is diminished. 
We find $d_0 -2R \sim (\mu-\mu_c(\alpha))^\nu$, where the average counterion 
concentration is $\mu=n/Z$ and the critical fraction $\mu_c$ depends on the valence 
of condensed counterions $\alpha$. From Fig.~\ref{d_0} it is evident that $\nu=1$. 
This should be contrasted with the line of charge model Ref.~\cite{Arenzon99}, 
for which $\nu=1/3$.  

\begin{figure}[t]
\vspace*{-.5cm}
\begin{center}
\epsfxsize=0.35\textwidth
\epsfbox{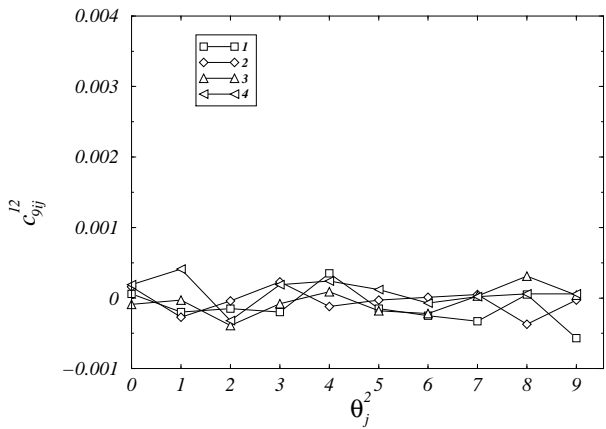}
\end{center}
\vspace*{-1cm}
\caption{Site-site correlation functions, Eq.~(\ref{correl}), for
the $i$'th site on the ring $z=9$ of the first polyion, with the $j$'th 
site on the $z=9$ ring of the second polyion. We consider only the 
correlations of the four ``inner sites'' of the first polyion $i=1,2,3,4$, 
(the four curves), with all the sites of the $z=9$ ring of the second 
polyion, $j=1...9$. The total number of sites per ring is $m=10$. $\theta_j^2$ 
indicates the angular position of the $j$'th site (in units of $2\pi /m$) 
on the ring of the second polyion. The parameters are as in Fig.~\ref{z20_force} 
with $n=7$ and the separation between the two macromolecules is $d/b=32.8$. 
The graph shows that at this distance there are almost no correlations 
between the condensed counterions. From Fig.~\ref{z20_force} we also see 
that the effective force is repulsive.}
\label{d32.8}
\end{figure}

\begin{figure}[t]
\vspace*{-.5cm}
\begin{center}
\epsfxsize=0.35\textwidth
\epsfbox{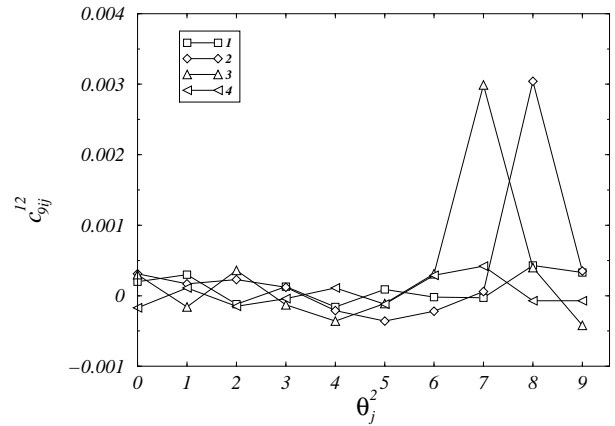}
\end{center}
\vspace*{-1cm}
\caption{Site-site correlation functions for separation $d/b=16.65$, where, 
according to Fig.~\ref{z20_force}, there is attraction. Note that sites 2 and 
3 (open diamond and triangle, respectively) on polyion 1 are strongly correlated 
with sites 8 and 7 on polyion 2, respectively}
\label{d16.65}
\end{figure}

In Fig.~\ref{snaps} we show two snapshots of the characteristic equilibrium 
configurations for (a) $d/b=32.8$ and (b) $d/b=16.8$. Looking at this figures 
it is difficult to see something that would distinguishes between them, both appear 
about the same. There is no obvious crystallization or transversal polarization 
suggested in previous studies~\cite{Niels97,Solis99}. Yet, the case (a) corresponds 
to the repulsive, while the case (b) corresponds to the attractive interaction between 
the polyions. To further explore this point, in Figs.~\ref{d32.8} and \ref{d16.65} 
we present the site-site correlation function, Eq.~(\ref{correl}), for macromolecules 
with $Z=20$ and $n=7$. For $d/b=32.8$ the surface-to-surface distance between the 
two polyions is sufficiently large for their condensed counterions to be practically 
uncorrelated, Fig.~\ref{d32.8}. On the other hand, for $d/b=16.65$ strong correlations 
between the condensed counterions are evident, Fig.~\ref{d16.65}. The Fig.~\ref{d16.65} 
shows that the sites two and three on the first polyion are strongly correlated 
with the sites seven and eight on the second polyion, respectively. It is these 
correlations between the adjacent sites on the two polyions which are responsible 
to the appearance of attraction between the two macromolecules when they are 
approximated, Fig.~\ref{z20_force}.           
 
\section{Summary}
\label{summary}

We have presented a simple model for polyion-polyion attraction inside a polyelectrolyte 
solution. It is clear from our calculations that the attraction results from the 
correlations between the condensed counterions and reaches maximum for $T=0$. 
The thermal fluctuations tend to diminish the correlations, decreasing the 
amplitude of the attractive force. Consistent with the experimental evidence, 
the attraction exists only in the presence of multivalent counterions. Our simulations 
demonstrate that a critical number of condensed counterions is necessary for the 
appearance of attraction.  The fraction of bare charge that must be neutralized for 
the attraction to arise depends on the valence of counterions. The larger the 
valence, the smaller the fraction of the bare polyion charger that must be neutralized 
for the attraction to appear.  This result should be contrasted with the line of 
charge model~\cite{Arenzon99} for which the critical fraction was found to be 
equal to $50\%$, independent of the counterion charge. 

\acknowledgments

We thank J. J. Arenzon for helpful comments on  simulations. This work was 
supported by CNPq --- Conselho Nacional de Desenvolvimento Cient{\'\i}fico e 
Tecnol{\'o}gico and FINEP --- Financiadora de Estudos e Projetos, Brazil.

\end{multicols}


\begin{thebibliography}{10}

\bibitem{Bloomfield91} V. A. Bloomfield, Biopolymers {\bf 31}, 1471 (1991).

\bibitem{Podgornik94} R. Podgornik, D. Rau, and V. A. Parsegian, Biophys. J. 
{\bf 66}, 962 (1994).

\bibitem{Israelachvili85} J. Israelachvili, {\it Intermolecular and Surface 
Forces} (Academic Press, London, 1985).

\bibitem{Rau92} D. C. Rau, and V. A. Parsegian, Biophys. J. {\bf 61}, 260 (1992).

\bibitem{Duguid95} J. G. Duguid, V. A. Bloomfield, J. M. Benevides, and 
G. J. Thomas, Jr., Biophys. J. {\bf 69}, 2633 (1995).

\bibitem{Knoll98} D. A. Knoll, M. D. Fried, and V. A. Bloomfield, in 
{\it DNA and its Drug Complexes}, edited by R. H. Sarma and M. H. Sarma 
(Adenin Press, New York, 1998). 

\bibitem{ras98} E. Raspaud, M. Olvera de la Cruz, J.-L. Sikorav, and
F. Livolant, Biophys. J. {\bf 74}, 381 (1998).


\bibitem{Tang96} J. X. Tang, S. Wong, P. Tran, and P. Janmey, Ber. Bunsen-Ges. 
Phys. Chem. {\bf 100}, 1 (1996).

\bibitem{Ray94} J. Ray and G.S. Manning, Langmuir {\bf 10}, 2450 (1994).

\bibitem{Ha97} B. -Y. Ha, and A. J. Liu, Phys. Rev. Lett. {\bf 79}, 1289 (1997).
 
\bibitem{Ha00} B. -Y. Ha, and A. J. Liu, {\it Physical Questions Posed by DNA 
Condensation}, to appear in {\it Physical Chemistry of Polyelectrolytes}, ed. 
T. Radeva (Marcel Dekker, New York, 2000).

\bibitem{Shklovsk199} I. Rouzina, V. Bloomfield, J. Chem. Phys. {\bf 100},
9977 (1996); B. I. Shklovskii, Phys. Rev. Lett. {\bf 82}, 3268 (1999)

\bibitem{Shklovsk299} B. I. Shklovskii, Phys. Rev. E {\bf 60}, 5802 (1999).

\bibitem{Perel99} V.I. Perel and B.I. Shklovskii, Physica A {\bf 274}, 5802 
(1999).

\bibitem{Nguyen00} T. T. Nguyen, I. Rouzina, and B. I. Shklovskii, J. Chem. 
Phys. {\bf 112}, 2562 (2000).

\bibitem{Levin99} Y. Levin, J. J. Arenzon, and J. F. Stilck, Phys. Rev. Lett. {\bf 83}, 
2680 (1999).

\bibitem{Arenzon99} J. J. Arenzon, J. F. Stilck, and Y. Levin, Euro. Phys. J. B. {\bf 12}, 
79 (1999).

\bibitem{Arenzon00} J. J. Arenzon, Y. Levin, and J. F. Stilck, Physica A {\bf 283}, 1 (2000).

\bibitem{Niels97} N. Gronbech-Jensen, R. J. Mashl, R. F. Bruisma, and 
M. W. Gelbart, Phys. Rev. Lett. {\bf 78}, 2577 (1997).

\bibitem{Solis99} F. J. Solis, and M. Olvera de la Cruz, Phys. Rev. E {\bf 60}, 4496 (1999).

\bibitem{Leikin99} A. A. Kornyshev and S. Leikin, Phys. Rev. Lett. {\bf 82}, 4138 (1999).

\bibitem{Lowen00} E. Allahyarov, and H. L\"owen, Phys. Rev. E {\bf 62}, 5542 (2000).

\bibitem{Allen} M. P. Allen, and D. J. Tildesley, {\it Computer Simulations of 
Liquids} (Clarendon, Oxford, 1987).

\bibitem{man69} G. S. Manning, J. Chem. Phys. {\bf 51}, 924 (1969); Q. Rev. Biophys.
{\bf 11}, 179 (1978).

\end{thebibliography}
\end{document}